\documentclass[aps,prl,twocolumn,showpacs]{revtex4-1}

\usepackage{bm,graphicx,textcomp,amssymb,amsmath,dcolumn,color}

\renewcommand{\Im}{\operatorname{Im}}

\newcommand{\ket}[1]{\left|#1\right>}
\newcommand{\bra}[1]{\left<#1\right|}

\newcommand{\nn}{\nonumber\\}

\newcommand{\f}[1]{\mbox{\boldmath$#1$}}
\newcommand{\fk}[1]{\mbox{\boldmath$\scriptstyle#1$}}
\newcommand{\vau}{\mbox{\boldmath$v$}}
\newcommand{\na}{\mbox{\boldmath$\nabla$}}
\newcommand{\bea}{\begin{eqnarray}}
\newcommand{\ea}{\end{eqnarray}}
\newcommand{\eea}{\end{eqnarray}}
\newcommand{\ord}{\,{\cal O}}

\begin{document}

\title{Reply to comment on ``Interaction of a BEC with a gravitational wave''}

\author{Ralf~Sch\"utzhold}

\affiliation{Helmholtz-Zentrum Dresden-Rossendorf, 
Bautzner Landstra{\ss}e 400, 01328 Dresden, Germany,}

\affiliation{Institut f\"ur Theoretische Physik, 
Technische Universit\"at Dresden, 01062 Dresden, Germany.}

\date{\today}

\begin{abstract}
This reply contains a brief response to the comment by R.~Howl, D.~R\"atzel, and I.~Fuentes. 
\end{abstract}

\maketitle

In case of conflicting points of view, a proper scientific debate can be very useful since it can 
help to address an issue from several angles and to identify the core of a problem.
Thus, let me respond to the comment \cite{comment} on my publication \cite{rs}. 
First, I would like to stress that my work in \cite{rs} is not meant to be a suggestion of a 
special (e.g., alternative to \cite{Sabin}) detection scheme, 
but rather a general investigation of the interaction 
between gravitational waves and Bose-Einstein condensates (BEC)
together with a general 
order-of-magnitude estimate of the interaction effects (as already stated in \cite{rs}).
In the comment \cite{comment}, I did not find any argument which invalidates 
the main results of \cite{rs}. 
Instead, I found several points of agreement -- so it might be best to start with those. 

As is well-known, for non-classical states (such as squeezed states or NOON states),
the sensitivity can scale with the number of particles (e.g., photons, atoms, or phonons) 
instead of its square root.
There also seems to be agreement that inhomogeneities in the condensate  
(e.g., weak inhomogeneities such as sound waves or strong inhomogeneities such as vortices) 
can enhance its interaction with gravitational waves. 
Furthermore, we seem to agree %there seems to be agreement 
that the number of %(linear) 
phonons (as linear excitations of the dilute BEC)
%in the (dilute) condensate 
should not exceed the number of atoms 
(even though the squeezing parameter $r=10$ proposed in \cite{Sabin} appears to suggest 
otherwise because it would correspond to a mean phonon number of $\sinh^2r>10^8$). 

Now, if the phonon number is well below the atom number, we may estimate the scaling of 
any perturbation Hamiltonian $\hat H_{\rm int}$ describing the interaction with a gravitational wave 
(including that stemming from the trap potential) which is bi-linear in the field operators $\hat\Psi$ 
by inserting the mean-field approximation $\hat\Psi\approx\psi_{\rm cond}+\hat\chi$,
where $\psi_{\rm cond}$ is the condensate wave-function and $\hat\chi$ describes the 
Bogoliubov-de~Gennes (phonon) modes.
We find the following hierarchy \cite{rs} of contributions: 
To leading order, the action of the perturbation Hamiltonian $\hat H_{\rm int}$
scales with the atom number.
To sub-leading order, it scales with the geometric mean of the atom and phonon number 
(or, if no phonons are present, with the square root of the atom number).
In this sequence, the effect considered in \cite{Sabin} would be next-to-sub-leading 
order as it scales with the number of phonons. 
Again, I found no argument in \cite{comment} which invalidates this picture. 

It is correct that the response of the trap potential to the gravitational wave is 
(apart from general estimates as sketched above) 
not considered explicitly in \cite{rs}. 
This would require a microscopic analysis of the trapping mechanism and how it 
changes due to the gravitational wave.
However, this is also not done in \cite{Sabin}, which is based on the %rather 
simplified assumption of a rigid and uniform box trap potential. 
Also, the direct creation of phonons by the trap 
(instead of the amplification of already existing phonons, 
see also \cite{Raetzel}) is not considered in \cite{Sabin}. 

Since the number of atoms or phonons in present-day BEC 
%Bose-Einstein condensates 
made of ultra-cold atomic vapor are not sufficient to reach the sensitivity required for 
gravitational wave detection, further large numbers are necessary.
This could be large ratios of length or time scales (or a large number of detectors). 
%(even though this might also be sorted into one of the above categories). 
%
The associated challenges %difficulties 
can be exemplarily illustrated by considering 
Eq.~(1) in \cite{comment}
\bea
\label{gl}
\Delta\epsilon\propto\frac{1}{N\sqrt{\omega_m\omega_n\tau tN_d}}
\,,
\ea
where $\Delta\epsilon$ measures the sensitivity, $N$ is the phonon number 
(denoted by $n$ in \cite{rs}), $\omega_m$ and $\omega_n$ are phonon frequencies,
$t$ is the run-time, $\tau$ the integration time, and $N_d$ denotes the 
number of detectors. 
Unfortunately, the authors of \cite{comment} did not provide explicit numbers, 
so let us insert example values suggested in the proposal \cite{Sabin}. 
Using the atom number of $10^6$ as assumed in \cite{Sabin}
as an upper bound for the phonon number $N$ (see above), 
the term within the square root in~\eqref{gl} should exceed $10^{30}$ 
in order to reach a sensitivity of $10^{-21}$, cf.~\cite{LIGO-paper}.
%
%(corresponding to typical gravitational waves, which are 
%required for gravitational wave detection, 
%which corresponds to the strongest signal measured at LIGO. 
%
As suggested in \cite{Sabin}, let us insert comparably large phonon frequencies 
in the $2\pi\times5~\rm kHz$ regime 
%%%%%%%%%%%%%%%%%%%%%%%%%%%%%%%%%%%%%%%
and a rather long run-time $t$ of 1000~seconds 
(with all the problems, see also \cite{rs,Raetzel}).
%
%The run-time $t$ is limited by the condensate life-time, the decoherence 
%time of the phonons and their Q-factor, the frequency stability of the 
%gravitational wave etc. 
%
%Let us also follow \cite{Sabin} and insert a rather long time of 1000~seconds
%(see also \cite{Raetzel}).
%
Still, we obtain an accumulated integration time $N_d\tau$ exceeding 
the age of our Universe, which seems to require an astronomical number 
$N_d$ of detectors/condensates. 

In summary, even though it would be extremely nice to detect gravitational 
waves with small present-day Bose-Einstein condensates made of ultra-cold 
atomic vapor, it appears to be a tremendously challenging task. 

%%%%%%%%%%%%%%%%%%%%%%%%%%%%%%%%%%%%%%%%%%%%%%%%%%%%%%%%%%%%%%%%%%%%%%%%%%%%%

\end{document}